\documentclass[10pt,conference,table]{IEEEtran}
\usepackage[switch]{lineno}
\IEEEoverridecommandlockouts

\usepackage{hyperref}
\usepackage{amsmath,amssymb,amsfonts}
\usepackage{algorithmic}
\usepackage[ruled,linesnumbered,noend]{algorithm2e}
\usepackage{xcolor}
\usepackage{graphicx}
\usepackage{textcomp}
\usepackage{xcolor}
\usepackage{nccmath}
\usepackage{environ}
\usepackage{tcolorbox}
\usepackage{soul}
\usepackage{booktabs}
\usepackage{multirow}
\usepackage{array}
\usepackage{float}
\usepackage{makecell}
\usepackage{flushend}
\usepackage{enumitem}
\usepackage{hhline}
\usepackage{comment}
\usepackage{epstopdf} 
\usepackage{caption}

\makeatletter
\newcommand{\removelatexerror}{\let\@latex@error\@gobble}
\makeatother

\newcommand{\code}[1]{\textnormal{\texttt{#1}}}

\begin{document}

\title{DeepCVA: Automated Commit-level Vulnerability Assessment with Deep Multi-task Learning}

\author{\IEEEauthorblockN{Triet Huynh Minh Le\IEEEauthorrefmark{1},
David Hin\IEEEauthorrefmark{1}\IEEEauthorrefmark{2},
Roland Croft\IEEEauthorrefmark{1}\IEEEauthorrefmark{2} and
M. Ali Babar\IEEEauthorrefmark{1}\IEEEauthorrefmark{2}}
\IEEEauthorblockA{\IEEEauthorrefmark{1}CREST - The Centre for Research on Engineering Software Technologies, The University of Adelaide, Australia}
\IEEEauthorblockA{\IEEEauthorrefmark{2}Cyber Security Cooperative Research Centre, Australia}
\IEEEauthorblockA{\{triet.h.le, david.hin, roland.croft, ali.babar\}@adelaide.edu.au}
}

\maketitle

\begin{abstract}
It is increasingly suggested to identify Software Vulnerabilities (SVs) in code commits to give early warnings about potential security risks. However, there is a lack of effort to assess vulnerability-contributing commits right after they are detected to provide timely information about the exploitability, impact and severity of SVs. Such information is important to plan and prioritize the mitigation for the identified SVs. We propose a novel Deep multi-task learning model, DeepCVA, to automate seven Commit-level Vulnerability Assessment tasks simultaneously based on Common Vulnerability Scoring System (CVSS) metrics. We conduct large-scale experiments on 1,229 vulnerability-contributing commits containing 542 different SVs in 246 real-world software projects to evaluate the effectiveness and efficiency of our model. We show that DeepCVA is the best-performing model with 38\% to 59.8\% higher Matthews Correlation Coefficient than many supervised and unsupervised baseline models. DeepCVA also requires 6.3 times less training and validation time than seven cumulative assessment models, leading to significantly less model maintenance cost as well. Overall, DeepCVA presents the first effective and efficient solution to automatically assess SVs early in software systems.
\end{abstract}


\section{Introduction}
Software Vulnerabilities (SVs) are security weaknesses that can make systems susceptible to cyber-attacks; thus, it is critical to assess SVs~\mbox{\cite{khan2018review}}. SV assessment is a process of determining characteristics of SVs such as attack vectors and impacts to help practitioners prioritize remediation for ever-increasing SVs~\mbox{\cite{smyth2017software}}. For example, SVs with simple exploitation and severe impacts likely require high fixing priority.

The expert-based Common Vulnerability Scoring System (CVSS) \mbox{\cite{cvss_website}} is a commonly used SV assessment framework. CVSS provides metrics to quantify exploitability, impact and severity level of SVs. However, there is usually delay in the manual process of assigning CVSS metrics to new SVs conducted by security experts~\mbox{\cite{feutrill2018effect}}.
Hence, there is an apparent need for automation in assessing reported/detected SVs.

Existing techniques (e.g.,~\mbox{\cite{lamkanfi2010predicting,han2017learning,spanos2018multi,le2019automated,le2021survey}}) to automate bug/SV assessment have mainly operated on bug/SV reports, but these reports may be only available long after SVs appeared in practice.
Our motivating analysis revealed that there were 1,165 days, on average, from when an SV was injected in a codebase until its report was published on National Vulnerability Database (NVD)~\cite{nvd_website}. Our analysis agreed with the findings of Meneely et al.~\mbox{\cite{meneely2013patch}}.
To tackle late-detected bugs/SVs, recently, Just-in-Time (commit-level) approaches (e.g.,~\mbox{\cite{hoang2019deepjit, kamei2012large,perl2015vccfinder,yang2017vuldigger}}) have been proposed to rely on the changes in code commits to detect bugs/SVs right after bugs/SVs are added to a codebase. Such early commit-level SV detection can also help reduce the delay in SV assessment.

Even when SVs are detected early in commits, we argue that existing automated techniques relying on bug/SV reports still struggle to perform \textit{just-in-time} SV assessment. Firstly, there are significant delays in the availability of SV reports, which render the existing SV assessment techniques unusable. Specifically, SV reports on NVD generally only appear seven days after the SVs are found/disclosed~\mbox{\cite{rodriguez2018analysis}}. Some of the detected SVs may not even be reported on NVD~\mbox{\cite{sawadogo2020learning}}, e.g., because of no disclosure policy. User-submitted bug/SV reports are also only available post-release and more than 82\% of the reports are filed more than 30 days after developers detected the bugs/SVs~\mbox{\cite{thung2012would}}.
Secondly, code review can provide faster SV assessment, but there are still unavoidable delays (from several hours to even days)~\mbox{\cite{bosu2012peer}}.
Delays usually come from code reviewers' late responses and manual analyses depending on the reviewers' workload and code change complexity~\mbox{\cite{thongtanunam2015investigating}}. Thirdly, it is non-trivial to automatically generate bug/SV reports from vulnerable commits as it would require non-code artefacts (e.g., stack traces or program crashes) that are mostly unavailable when commits are submitted~\mbox{\cite{lamkanfi2010predicting,moran2015auto}}.

Performing commit-level SV assessment provides a possibility to inform committers about the exploitability, impact and severity of SVs in code changes and prioritize fixing earlier \textit{without waiting for SV reports}. However, to the best of our knowledge, there is no existing work on automating SV assessment in commits. Prior SV assessment techniques that analyze text in SV databases (e.g.,~\mbox{\cite{han2017learning,spanos2018multi,le2019automated}}) also cannot be directly adapted to the commit level. Contrary to text, commits contain deletions and additions of code with specific structure and semantics~\mbox{\cite{hoang2019deepjit,hoang2019patchnet}}. Additionally, we speculate that CVSS metrics can be related. For example, an SQL injection is likely to be highly severe since attackers can exploit it easily via crafted input and compromise data confidentiality and integrity. We posit that these metrics would have common patterns in commits that can be potentially shared between SV assessment models. Predicting related tasks in a shared model has been successfully utilized for various applications~\mbox{\cite{zhang2017survey}}. For instance, an autonomous car is driven with simultaneous detection of vehicles, lanes, signs and pavement~\mbox{\cite{chowdhuri2019multinet}}. These observations motivated us to tackle a new and important research challenge, \textbf{``How can we leverage the common attributes of assessment tasks to perform effective and efficient commit-level SV assessment?''}

We present DeepCVA, a novel \mbox{\underline{\textbf{Deep}}} multi-task learning model, to automate \mbox{\underline{\textbf{C}}}ommit-level \mbox{\underline{\textbf{V}}}ulnerability \mbox{\underline{\textbf{A}}}ssessment. DeepCVA first uses attention-based convolutional gated recurrent units to extract features of code and surrounding context from vulnerability-contributing commits (i.e., commits with vulnerable changes). The model uses these features to predict seven CVSS assessment metrics (i.e., Confidentiality, Integrity, Availability, Access Vector, Access Complexity, Authentication, and Severity) simultaneously using the multi-task learning paradigm. The predicted CVSS metrics can guide SV management and remediation processes.

Our key contributions are summarized as follows:

\begin{itemize}[noitemsep,topsep=0pt]     
    \item We are the first to tackle the commit-level SV assessment tasks that enable early security risks estimation and planning for SV remediation.
    \item We propose a unified model, DeepCVA, to automate seven commit-level SV assessment tasks simultaneously.
    \item We extensively evaluate DeepCVA on our curated large-scale dataset of 1,229 vulnerability-contributing commits with 542 SVs from 246 real-world projects.
    \item We demonstrate that DeepCVA has 38\% to 59.8\% higher Matthews Correlation Coefficient (MCC) than various supervised and unsupervised baseline models using text-based features and software metrics. The proposed context-aware features improve the MCC of DeepCVA by 14.8\%. The feature extractor with attention-based convolutional gated recurrent units, on average, adds 52.9\% MCC for DeepCVA. Multi-task learning also makes DeepCVA 24.4\% more effective and 6.3 times more efficient in training/validation/testing than separate models for seven assessment tasks.
    \item We release our source code, models and datasets at~\mbox{\cite{reproduction_package_ase2021}}.
\end{itemize}

\noindent \textbf{Paper structure}. Section~\mbox{\ref{sec:background}} introduces preliminaries and motivation. Section~\mbox{\ref{sec:deepcva_method}} proposes the DeepCVA model for commit-level SV assessment. Section~\mbox{\ref{sec:expt_design}} describes our experimental design and setup. Section~\mbox{\ref{sec:results}} presents the experimental results. Section~\mbox{\ref{sec:discussion}} discusses our findings and threats to validity. Section~\mbox{\ref{sec:related_work}} covers the related work. Section~\mbox{\ref{sec:conclusions}} concludes the work and proposes future directions.

\section{Background and Motivation}
\label{sec:background}
\subsection{Vulnerability in Code Commits}
\label{subsec:vuln_commits}

\begin{figure}[t]
    \centering
    \includegraphics[width=\columnwidth,keepaspectratio]{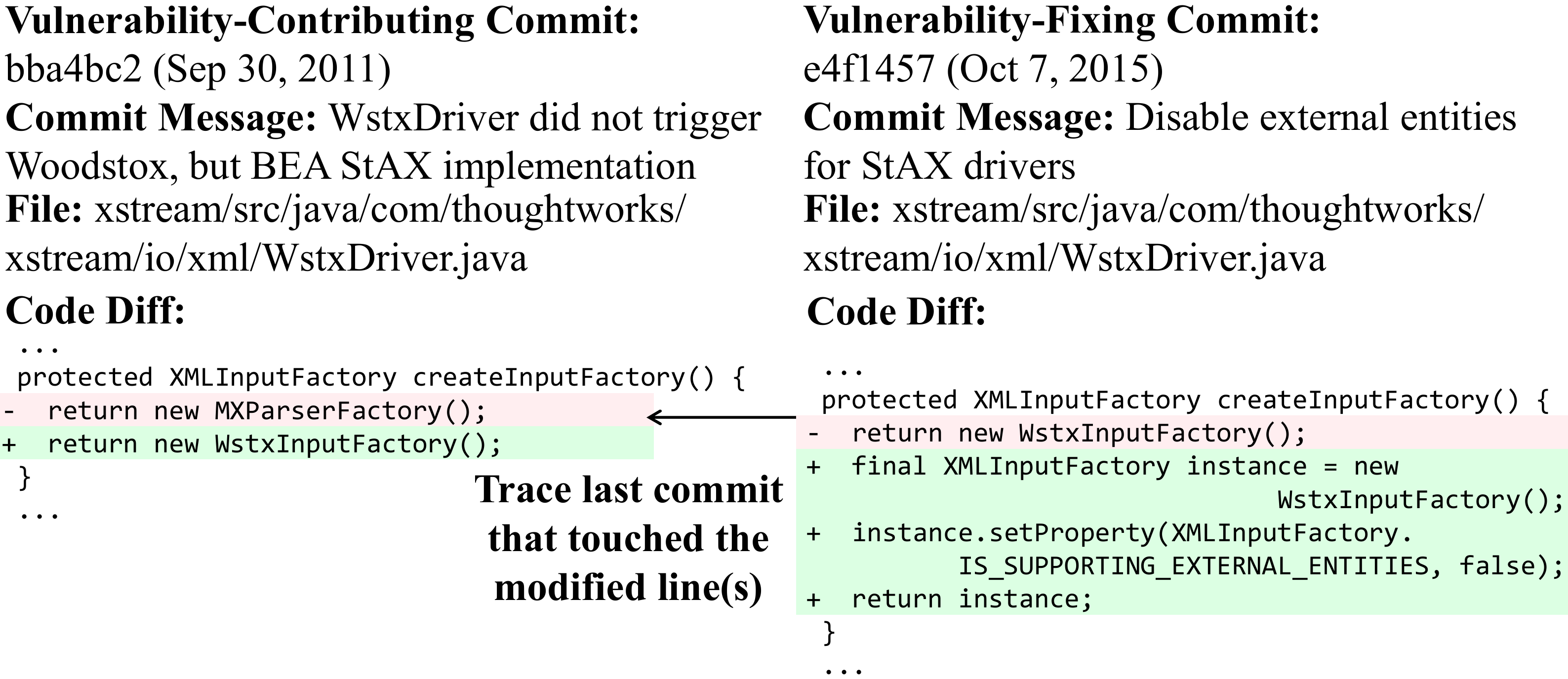}
    \caption{Exemplary SV fixing commit (right) for the XML external entity injection (XXE) (CVE-2016-3674) and its respective SV contributing commit (left) in the \textit{xstream} project.}
    \vspace{-9pt}
    \label{fig:commit_ex}
\end{figure}

Commits are an essential unit of any version control system (e.g., Git) and record all the chronological changes made to the codebase of a software project. As illustrated in Fig.~\mbox{\ref{fig:commit_ex}}, changes in a commit consist of deletion(s) (–) and/or addition(s) (+) in each affected file.

Vulnerability-Contributing Commits (VCCs) are commits whose changes contain SVs~\mbox{\cite{meneely2013patch}}, e.g., using vulnerable libraries or insecure implementation. We focus on VCCs rather than any commits with vulnerable code (in unchanged parts) since addressing VCCs helps mitigate SVs as early as they are added to a project.
VCCs are usually obtained based on Vulnerability-Fixing Commits (VFCs)~\mbox{\cite{perl2015vccfinder, yang2017vuldigger}}. An exemplary VFC and its respective VCC are shown in Fig.~\mbox{\ref{fig:commit_ex}}. VFCs delete, modify or add code to eliminate an SV (e.g., disabling external entities processing in the XML library in Fig.~\mbox{\ref{fig:commit_ex}}) and can be found in bug/SV tracking systems. Then, VCCs are commits that last touched the code changes in VFCs.
Our work also leverages VFCs to obtain VCCs for building automated commit-level SV assessment models.

\begin{figure*}[t]
    \centering
    \includegraphics[width=0.99\textwidth,keepaspectratio]{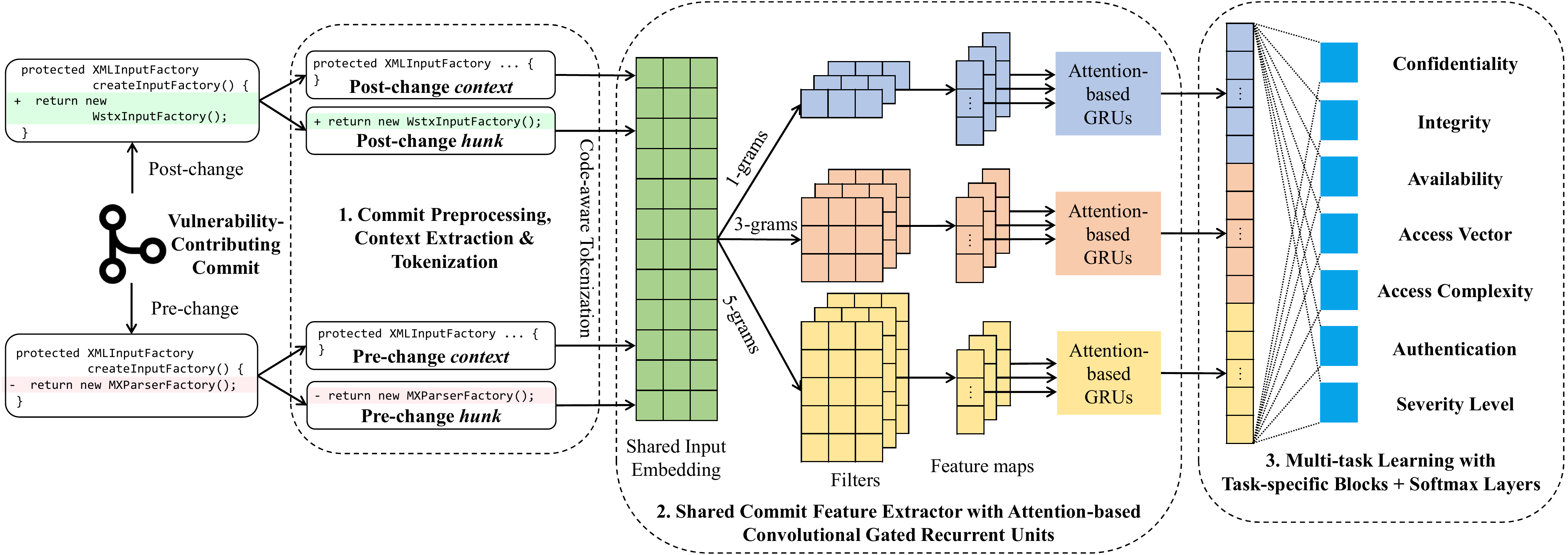}
    \caption{Workflow of DeepCVA for automated commit-level SV assessment. The VCC is the one described in Fig.~\mbox{\ref{fig:commit_ex}}.}
    \label{fig:workflow_deepcva}
\end{figure*}

\subsection{Commit-level SV Assessment with CVSS}
\label{subsec:cva_need}

Common Vulnerability Scoring System (CVSS)~\mbox{\cite{cvss_website}} has been an expert-maintained standard for SV assessment. CVSS base metrics are prevalently used to determine through which attack vectors SVs can be exploited and assess their potential impacts. This allows developers to better plan and prioritize the mitigation of such SVs. The base metrics are \textit{Confidentiality}, \textit{Integrity}, \textit{Availability}, \textit{Access Vector}, \textit{Access Complexity}, \textit{Authentication} and \textit{Severity}. We use CVSS version 2 of base metrics to assess SVs as version 2 is still more predominantly used than version 3 (introduced in 2015). SVs before 2015 are also still relevant in the modern context; e.g., CVE-2004-0113 discovered in 2004 was exploited in a crypto attack in 2018~\mbox{\cite{old_sv_exploit}}. Based on CVSS version 2, the VCC (CVE-2016-3674) in Fig.~\mbox{\ref{fig:commit_ex}} has a considerable impact on the Confidentiality. This SV can be exploited with low (Access) complexity with no authentication via public network (Access Vector), making it an attractive target for attackers.

Despite the criticality of these SVs, there have been delays in reporting, assessing and fixing them. Concretely, the VCC in Fig.~\mbox{\ref{fig:commit_ex}} required 1,439 and 1,469 days to be reported\footnote{https://github.com/x-stream/xstream/issues/25} and fixed (in VFC), respectively. Existing SV assessment methods based on bug/SV reports (e.g.,~\mbox{\cite{han2017learning,spanos2018multi,le2019automated}}) would need to wait more than 1,000 days for the report of this SV. However, performing SV assessment right after this commit was submitted can bypass the waiting time for SV reports, enabling developers to realize the exploitability/impacts of this SV and plan to fix it much sooner.
To the best of our knowledge, there has not been any study addressing automated commit-level SV assessment, i.e., assigning seven CVSS base metrics to a VCC. Our work identifies and aims to bridge this important research gap.

\subsection{Feature Extraction from Commit Code Changes}
\label{subsec:sv_context_motivation}

The extraction of commit features is important for building commit-level SV assessment models.
Many existing commit-level defect/SV prediction models have only considered commit code changes (e.g.,~\mbox{\cite{hoang2019deepjit,hoang2019patchnet,sabetta2018practical}}). However, we argue that the nearby context of code changes also contributes valuable information to the prediction. For instance, the surrounding code of the changes in Fig.~\mbox{\ref{fig:commit_ex}} provides extra details; e.g., the method return statement is modified and the return type is \code{XMLInputFactory}. Such a type can help learn properties of XXE SV that usually occurs with XML processing.

Besides the context, we speculate that SV assessment models can also benefit from the relatedness among the assessment tasks. For example, the XXE SV in Fig.~\mbox{\ref{fig:commit_ex}} allows attackers to read arbitrary system files, which mainly affects the Confidentiality rather than the Integrity and Availability of a system. This work investigates the possibility of incorporating the common features of seven CVSS metrics into a single model using the multi-task learning paradigm~\mbox{\cite{zhang2017survey}} instead of learning seven cumulative individual models. Specifically, multi-task learning leverages the similarities and the interactions of the involved tasks through a shared feature extractor to predict all the tasks simultaneously. Such a unified model can significantly reduce the time and resources to train, optimize and maintain/update the model in the long run.

\section{The DeepCVA Model}
\label{sec:deepcva_method}

We propose \textbf{DeepCVA} (see Fig.~\mbox{\ref{fig:workflow_deepcva}}), a novel \underline{\textbf{Deep}} learning model to automate \underline{\textbf{C}}ommit-level \underline{\textbf{V}}ulnerability \underline{\textbf{A}}ssessment. DeepCVA is a unified and end-to-end trainable model that concurrently predicts seven CVSS metrics (i.e., Confidentiality, Integrity, Availability, Access Vector, Access Complexity, Authentication, and Severity) for a Vulnerability-Contributing Commit (VCC).
DeepCVA contains: (\textit{i}) preprocessing, context extraction and tokenization of code commits (section~\mbox{\ref{subsec:preprocessing}}), (\textit{ii}) feature extraction from commits shared by seven assessment tasks using attention-based convolutional gated recurrent units (section~\mbox{\ref{subsec:deep_acgru}}), and (\textit{iii}) simultaneous prediction of seven CVSS metrics using multi-task learning~\mbox{\cite{zhang2017survey}} (section~\mbox{\ref{subsec:multitask_learning}}).
To assign the CVSS metrics to a new VCC with DeepCVA, we first preprocess the commit, obtain its code changes and respective context and tokenize such code changes/context. Embedding vectors of preprocessed code tokens are then obtained, and the commit feature vector is extracted using the trained feature extractor. This commit feature vector passes through the task-specific blocks and softmax layers to get the seven CVSS outputs with the highest probability values.
Details of each component are given hereafter.

\subsection{Commit Preprocessing, Context Extraction \& Tokenization}
\label{subsec:preprocessing}

To train DeepCVA, we first obtain and preprocess code changes (hunks) and extract the context of such changes. We then tokenize them to prepare inputs for feature extraction.

\noindent \textbf{Commit preprocessing}. Preprocessing helps remove noise in code changes and reduce computational costs. We remove newlines/spaces and inline/multi-line comments since they do not change code functionality. We do not remove punctuations (e.g., ``\code{;}'', ``\code{(}'', ``\code{)}'') and stop words (e.g., \code{and}/\code{or} operators) to preserve code syntax. We also do not lowercase code tokens since developers can use case-sensitivity for naming conventions of different token types (e.g., variable name: \code{system} vs. class name: \code{System}). Stemming (i.e., reducing a word to its root form such as \code{equals} to \code{equal}) is not applied to code since different names can change code functionality (e.g., the built-in \code{equals} function in Java).

\begin{figure}[t]
    \centering
    \includegraphics[width=0.99\columnwidth,keepaspectratio]{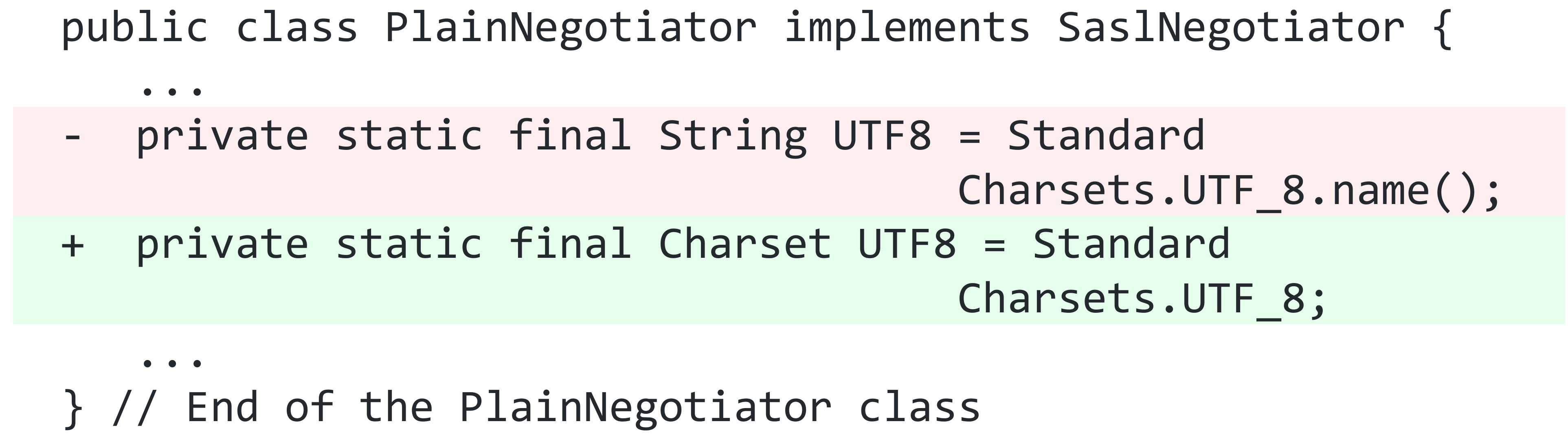}
    \caption{Code changes outside of a method from the commit \textit{4b9fb37} in the \textit{Apache qpid-broker-j} project.}
    \label{fig:method_scope}
\end{figure}

\noindent \textbf{Context extraction algorithm}. We customize Sahal et al.'s~\mbox{\cite{sahal2018identifying}} \textit{Closest Enclosing Scope} (CES) to identify the context of vulnerable code changes for commit-level SV assessment (see section~\mbox{\ref{subsec:sv_context_motivation}}). Sahal et al.~\mbox{\cite{sahal2018identifying}} defined an enclosing scope to be the code within a balanced amount of opening and closing curly brackets such as \code{if}/\code{switch}/\code{while}/\code{for} blocks.
Among all enclosing scopes of a hunk, the one with the smallest size (lines of code) is selected as CES to reduce irrelevant code. Sahal et al.~\mbox{\cite{sahal2018identifying}} found CES usually contains hunk-related information (e.g., variable values/types preceding changes). CES also alleviates the need for manually pre-defining the context size as in~\mbox{\cite{perl2015vccfinder,tian2020evaluating}}.
Some existing studies (e.g.,~\mbox{\cite{li2020dlfix,alon2019code2vec}}) only used the method/function scope, but code changes may occur outside of a method. For instance, changes in Fig.~\mbox{\ref{fig:method_scope}} do not have any enclosing method, but we can still obtain its CES, i.e., the \code{PlainNegotiator} class.

There are still two main limitations with the definition of CES in~\mbox{\cite{sahal2018identifying}}. Firstly, a scope (e.g., \code{for}/\code{while} in Java) with single-line content does not always require curly brackets. Secondly, some programming languages do not use curly brackets to define scopes like Python. To address these two issues, we utilize Abstract Syntax Tree (AST) depth-first traversal (see Algorithm~\mbox{\ref{algo:context_extraction}}) to obtain CESs of code changes, as AST covers the syntax of all scope types and generalizes to any programming languages.

Algorithm~\mbox{\ref{algo:context_extraction}} contains: (\textit{i}) the \code{extract\_scope} function for extracting potential scopes of a code hunk (lines 1-8), and (\textit{ii}) the main code to obtain the CES of every hunk in a commit (lines 9-18). The \code{extract\_scope} function leverages depth-first traversal with recursion to go through every node in an AST of a file. Line 3 adds the selected part of an AST to the list of potential scopes (\code{potential\_scopes}) of the current hunk. The first (root) AST is always valid since it encompasses the whole file. Line 6 then checks whether each node (sub-tree) of the current AST has one of the following types: \code{class}, \code{interface}, \code{enum}, \code{method}, \code{if}/\code{else}, \code{switch}, \code{for}/\code{while}/\code{do}, \code{try}/\code{catch}, and is surrounding the current hunk. If the conditions are satisfied, the \code{extract\_scope} function would be called recursively in line 7 until a leaf of the AST is reached. The main code starts to extract the modified files of the current commit in line 9. For each file, we extract code hunks (code deletions/additions) in line 12 and then obtain the AST of the current file using an AST parser in line 13. Line 16 calls the defined \code{extract\_scope} function to generate the potential scopes for each hunk. Among the identified scopes, line 17 adds the one with the smallest size (i.e., the number of code lines excluding empty lines and comments) to the list of CESs (\code{all\_ces}). Finally, line 18 of Algorithm~\mbox{\ref{algo:context_extraction}} returns all the CESs for the current commit.

We treat deleted (pre-change), added (post-change) code changes and their CESs as four separate inputs to be vectorized by the shared input embedding, as illustrated in Fig.~\mbox{\ref{fig:workflow_deepcva}}. For each input, we concatenate all the hunks/CESs in all the affected files of a commit to explicitly capture their interactions.

\noindent \textbf{Code-aware tokenization}. The four inputs extracted from a commit are then tokenized with a code-aware tokenizer to preserve code semantics and help prediction models be more generalizable. For example, \code{a++} and \code{b++} are tokenized as \code{a}, \code{b} and \code{++}, explicitly giving a model the information about one-increment operator (\code{++}).
Tokenized code is fed into a shared Deep Learning model, namely Attention-based Convolutional Gated Recurrent Unit (AC-GRU), to extract commit features.

\setlength{\textfloatsep}{6pt}
\begin{algorithm}[t]
\fontsize{8}{9}\selectfont
    \caption{AST-based extraction of the Closest Enclosing Scopes (CESs) of commit code changes.}
    \label{algo:context_extraction}
    \DontPrintSemicolon
    \SetAlgoNoLine

    \KwIn{Current Vulnerability-Contributing Commit (VCC): $commit$\\
    Scope type: $scope\_types$}

    \KwOut{CESs of code changes in the current commit: $all\_ces$}
    
    \SetKwFunction{FMain}{extract\_scope}
    \SetKwProg{Fn}{Function}{:}{}
    \Fn{\FMain{$AST, hunk, visited=\emptyset$}}{
    
    \textbf{global} $potential\_scopes$\;
        
    $potential\_scopes \longleftarrow potential\_scopes + AST$\;
    
    $visited \longleftarrow visited + AST$\;
    
    \ForEach{$node \in AST$}{
        \If{$ node \notin visited$ {\bf and} $\text{type}(node) \in scope\_types~$ {\bf and} $start_{node}\leq start_{hunk}$ {\bf and} $end_{node} \geq end_{hunk}$}{
            {$extract\_scope(AST,hunk,visited)$}
        }
    }
    \textbf{return}\; 
    }

    $files \longleftarrow \text{extract\_files}(commit)$\;
    $all\_ces \longleftarrow \emptyset$\;

    \ForEach{${f}_{i}\in files$}{
        $hunks \longleftarrow \text{extract\_hunk}(commit,f_{i})$\;
        
        $AST_{i} \longleftarrow \text{extract\_AST}(f_{i})$\;
        
        \ForEach{${h}_{i}\in hunks$}{
            $potential\_scopes \longleftarrow \emptyset$\;
            
            $\text{extract\_scopes}(AST_{i}, h_{i})$\;
            
            $all\_ces \longleftarrow all\_ces + \underset{\text{size}}{\mathop{\operatorname{argmin}}}(potential\_scopes)$\;    
        }
    }

    \Return $all\_ces$
\end{algorithm}

\subsection{Feature Extraction with Deep AC-GRU}
\label{subsec:deep_acgru}

Deep AC-GRU has a \textit{three-way Convolutional Neural Network} to extract n-gram features and \textit{Attention-based Gated Recurrent Units} to capture dependencies among code changes and their context. This feature extractor is shared by four inputs, i.e., deleted/added code hunks/context. Each input has the size of $N \times L$, where $N$ is the no. of code tokens and $L$ is the vector length of each token. All inputs are truncated or padded to the same length $N$ to support parallelization. The feature vector of each input is obtained from a shared \textit{Input Embedding} layer that maps code tokens into fixed-length arithmetic vectors. The dimensions of this embedding layer are $|V| \times L$, where $|V|$ is the code vocabulary size, and its parameters are learned together with the rest of the model.

\noindent \textbf{Three-way Convolutional Neural Network}. We use a shared three-way Convolutional Neural Network (CNN)~\mbox{\cite{kim2014convolutional}} to extract n-grams (n = 1,3,5) of each input vector. The three-way CNN has filters with three sizes of one, three and five, respectively, to capture common code patterns, e.g., \code{public class Integer}.
The filters are randomly initialized and jointly learned with the other components of DeepCVA. We did not include 2-grams and 4-grams to reduce the required computational resources without compromising the model performance, which has been empirically demonstrated in section~\mbox{\ref{subsec:rq2_results}}. To generate code features of different window sizes with the three-way CNN, we multiply each filter with the corresponding input rows and apply non-linear ReLU activation function~\mbox{\cite{nair2010rectified}}, i.e., ReLU(x) = $\text{max}(0, x)$. We repeat the same convolutional process from the start to the end of an input vector by moving the filters down sequentially with a stride of one.
\textcolor{black}{This stride value is the smallest and helps capture the most fine-grained information from input code as compared to larger values.}
Each filter size returns feature maps of the size $(N - K + 1) \times F$, where $K$ is the filter size (one, three or five) and $F$ is the number of filters. Multiple filters are used to capture different semantics of commit data.

\noindent \textbf{Attention-based Gated Recurrent Unit}. The feature maps generated by the three-way CNN sequentially enter a Gated Recurrent Unit (GRU) [31]. GRU, defined in Eq.~\mbox{\eqref{eq:gru_formula}}, is an efficient version of Recurrent Neural Networks and used to explicitly capture the order and dependencies between code blocks. For example, the \code{return} statement comes after the function declarations of the VCC in Fig.~\mbox{\ref{fig:workflow_deepcva}}.

\begin{equation}\label{eq:gru_formula}
\begin{aligned}
  & {{\mathbf{z}}_{t}}=\sigma ({{\mathbf{W}}_{z}}{{\mathbf{x}}_{t}}+{{\mathbf{U}}_{z}}{{\textbf{h}}_{t-1}}+{{b}_{z}}) \\
 & {{\mathbf{r}}_{t}}=\sigma ({{\mathbf{W}}_{r}}{{\mathbf{x}}_{t}}+{{\mathbf{U}}_{r}}{{\mathbf{h}}_{t-1}}+{{b}_{r}}) \\
 & {{{\mathbf{\hat{h}}}}_{t}}=\tanh ({{\mathbf{W}}_{h}}{{\mathbf{x}}_{t}}+{{\mathbf{U}}_{h}}({{\mathbf{r}}_{t}}\odot {{\mathbf{h}}_{t-1}})+{{b}_{h}}) \\
 & {{\mathbf{h}}_{t}}=(1-{\mathbf{z}_{t}})\odot {{\mathbf{h}}_{t-1}}+{\mathbf{z}_{t}}\odot {{{\mathbf{\hat{h}}}}_{t}}
\end{aligned}
\end{equation}
\noindent where $\mathbf{W}_{z}$, $\mathbf{W}_{r}$, $\mathbf{W}_{h}$, $\mathbf{U}_{z}$, $\mathbf{U}_{r}$, $\mathbf{U}_{h}$ are learnable weights, $b_{z}$, $b_{r}$, $b_{h}$ are learnable biases, $\odot$ is element-wise multiplication, $\sigma$ is sigmoid function and $\tanh()$ is hyperbolic tangent function.

\noindent To determine the information ($\textbf{h}_{t}$) at each token (time step) $t$, GRU combines the current input ($\textbf{x}_{t}$) and the previous time step ($\textbf{h}_{t-1}$) using the \textit{update} ($\textbf{z}_{t}$) and \textit{reset} ($\textbf{r}_{t}$) gates. $\textbf{h}_{t}$ is then carried on to the next token until the end of the input to maintain the dependencies of the whole code sequence.

The last token output of GRU is often used as the whole sequence representation, yet it suffers the \textit{information bottleneck} problem~\mbox{\cite{bahdanau2014neural}}, especially for long sequences.
To address this issue, we incorporate the \textit{attention mechanism}~\mbox{\cite{bahdanau2014neural}} into GRU to explicitly capture the contribution of each input token, as formulated in Eq.~\mbox{\eqref{eq:attention}}.
\begin{equation}\label{eq:attention}
\begin{aligned}
  & \mathbf{ou}{{\mathbf{t}}_{attention}}=\sum\limits_{i=1}^{m}{{{w}_{i}}{{\mathbf{h}}_{i}}} \\ 
 & {{w}_{i}}=\operatorname{softmax}({{\mathbf{W}}_{s}}\tanh ({{\mathbf{W}}_{a}}{{\mathbf{h}}_{i}}+{{b}_{a}})) \\ 
 & =\frac{\exp ({{\mathbf{W}}_{s}}\tanh ({{\mathbf{W}}_{a}}{{\mathbf{h}}_{i}}+{{b}_{a}}))}{\sum\limits_{j=1}^{m}{\exp ({{\mathbf{W}}_{s}}\tanh ({{\mathbf{W}}_{a}}{{\mathbf{h}}_{j}}+{{b}_{a}}))}}
\end{aligned}
\end{equation}
\noindent where $w_{i}$ is the weight of $\mathbf{h}_{i}$; $\mathbf{W}_{s}$, $\mathbf{W}_{a}$ are learnable weights, $b_{a}$ is learnable bias, and $m$ is the number of code tokens.

The attention-based outputs ($\mathbf{out}_{attention}$) of the three GRUs (see Fig.~\mbox{\ref{fig:workflow_deepcva}}) are concatenated into a single feature vector to represent each of the four inputs (pre-/post-change hunks/contexts). The commit feature vector is a concatenation of the vectors of all four inputs generated by the shared AC-GRU feature extractor. This feature vector is used for multi-task prediction of seven CVSS metrics.

\subsection{Commit-level SV Assessment with Multi-task Learning}
\label{subsec:multitask_learning}

This section describes the multi-task learning layers of DeepCVA for efficient commit-level SV assessment using a single model as well as how to train the model end-to-end.

\noindent \textbf{Multi-task learning layers}. The last component of DeepCVA consists of the multi-task learning layers that simultaneously give the predicted CVSS values for seven SV assessment tasks. As illustrated in Fig.~\mbox{\ref{fig:workflow_deepcva}}, this component contains two main parts: \textit{task-specific blocks} and \textit{softmax layers}. On top of the shared features extracted by AC-GRU, task-specific blocks are necessary to capture the differences among the seven tasks. Each task-specific block is implemented using a fully connected layer with non-linear ReLU activations~\mbox{\cite{nair2010rectified}}. Specifically, the output vector ($\mathbf{task}_{i}$) of the task-specific block for assessment task $i$ is defined in Eq.~\mbox{\eqref{eq:task_specific_block}}.
\begin{equation}\label{eq:task_specific_block}
\mathbf{tas}{{\mathbf{k}}_{i}}=\operatorname{ReLU}({{\mathbf{W}}_{t}}{{\mathbf{x}}_{commit}}+{{b}_{t}})
\end{equation}
\noindent where $\mathbf{x}_{commit}$ is the commit feature vector from AC-GRU, $\mathbf{W}_{t}$ is learnable weights and $b_{t}$ is learnable bias.

Each task-specific vector goes through the respective softmax layer to determine the output of each task with the highest predicted probability. The prediction output ($pred_{i}$) of task $i$ is given in Eq.~\mbox{\eqref{eq:softmax_layer}}.
\begin{equation}\label{eq:softmax_layer}
\begin{aligned}
  & pre{{d}_{i}}=\operatorname{argmax}(\mathbf{pro}{{\mathbf{b}}_{i}}) \\ 
 & \mathbf{pro}{{\mathbf{b}}_{i}}=\operatorname{softmax}({{\mathbf{W}}_{p}}\mathbf{tas}{{\mathbf{k}}_{i}}+{{b}_{p}}) \\ 
 & \operatorname{softmax}({{z}_{j}})=\frac{\exp ({{z}_{j}})}{\sum\limits_{c=1}^{nlabel{{s}_{i}}}{\exp ({{z}_{c}})}}
\end{aligned}
\end{equation}
\noindent where $\mathbf{prob}_{i}$ contains the predicted probabilities of $nlabels_{i}$ possible outputs of task $i$; $\mathbf{W}_{p}$ is learnable weights and $b_{p}$ is learnable bias.

\noindent \textbf{Training DeepCVA}. To compare DeepCVA's outputs with ground-truth CVSS labels, we define a multi-task loss that averages the cross-entropy losses of seven tasks in Eq.~\mbox{\eqref{eq:multiloss}}.
\begin{equation}\label{eq:multiloss}
\begin{aligned}
  & los{{s}_{DeepCVA}}=\sum\limits_{i=1}^{7}{los{{s}_{i}}} \\ 
 & los{{s}_{i}}=-\sum\limits_{c=1}^{nlabel{{s}_{i}}}{y_{i}^{c}\log (prob_{i}^{c})},\,y_{i}^{c}=1\,\text{if}\,c\,\text{is}\,\text{true}\,\text{class}\,\text{else}\,0 \\ 
\end{aligned}
\end{equation}
\noindent where $y_{i}^{c}$, $prob_{i}^{c}$, and $nlabels_{i}$ are the ground-truth value, predicted probability and all labels of CVSS task $i$, respectively.

We minimize this multi-task loss using a stochastic gradient descent method~\mbox{\cite{ruder2016overview}} to optimize the weights of learnable components in DeepCVA. We also use backpropagation~\mbox{\cite{rumelhart1986learning}} to automate partial differentiation with chain-rule and increase the efficiency of gradient computation throughout the model.

\section{Experimental Design and Setup}
\label{sec:expt_design}

All the experiments ran on a computing cluster that has 16 CPU cores with 16GB of RAM and Tesla V100 GPU.

\subsection{Datasets}
\label{subsec:datasets}

To develop commit-level SV assessment models, we built a dataset of Vulnerability-Contributing Commits (VCCs) and their CVSS metrics. We used Vulnerability-Fixing Commits (VFCs) to retrieve VCCs, as discussed in section~\mbox{\ref{subsec:vuln_commits}}.

\noindent \textbf{VFC identification}. We first obtained VFCs from three public sources: NVD~\mbox{\cite{nvd_website}}, GitHub and its Advisory Database\footnote{https://github.com/advisories} as well as a manually curated/verified VFC dataset (VulasDB)~\mbox{\cite{ponta2019manually}}. In total, we gathered 13,310 VFCs that had dates ranging from July 2000 to October 2020. We selected VFCs in Java projects as Java has been commonly investigated in the literature (e.g.,~\mbox{\cite{hoang2019deepjit,alon2019code2vec,mcintosh2017fix}}) and also in the top five most popular languages in practice.\footnote{https://insights.stackoverflow.com/survey/2020\#technology-most-loved-dreaded-and-wanted-languages-loved}
Following the practice of~\mbox{\cite{mcintosh2017fix}}, we discarded VFCs that had more than 100 files and 10,000 lines of code to reduce noise in the data.

\begin{figure}[t]
    \centering
    \includegraphics[width=\columnwidth,keepaspectratio]{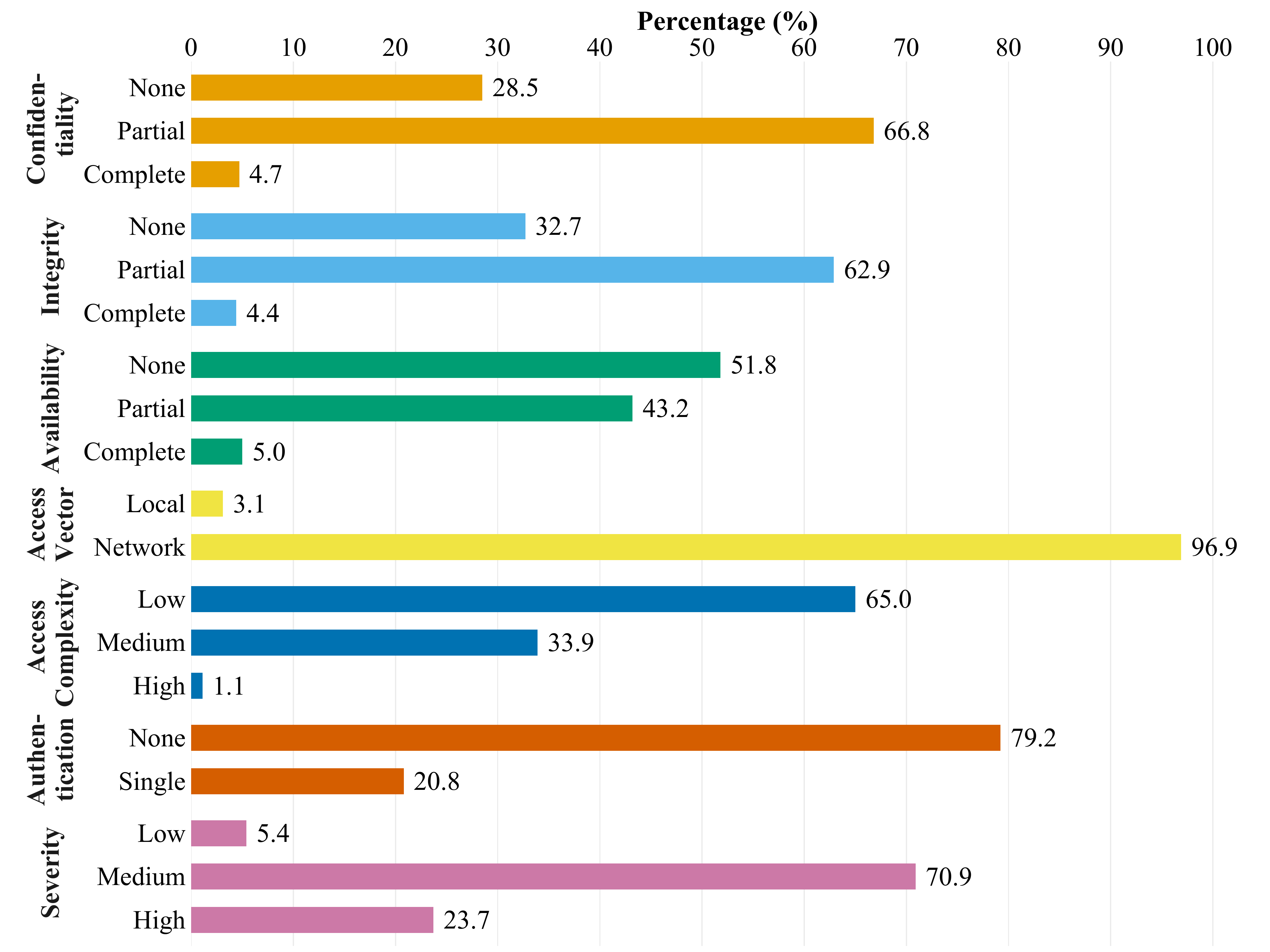}
    \caption{Data distributions of seven SV assessment tasks.}
    \label{fig:cvss_distribution}
\end{figure}

\noindent \textbf{VCC identification with the SZZ algorithm}. After the filtering steps, we had 1,602 remaining unique VFCs to identify VCCs using the SZZ algorithm~\mbox{\cite{sliwerski2005changes}}. This algorithm selects commits that last modified the source code lines deleted or modified to address an SV in a VFC as the respective VCCs of the same SV (see Fig.~\mbox{\ref{fig:commit_ex}}). As in~\mbox{\cite{sliwerski2005changes}}, we first discarded commits with timestamps after the published dates of the respective SVs on NVD since SVs can only be reported after they were injected in a codebase. We then removed cosmetic changes (e.g., newlines and white spaces) and single-line/multi-line comments in VFCs since these elements do not change code functionality~\mbox{\cite{mcintosh2017fix}}. Like~\mbox{\cite{mcintosh2017fix}}, we also considered copied or renamed files while tracing VCCs.
We obtained 1,229 unique VCCs\footnote{The SV reports of all curated VCCs were not available at commit time.} of 542 SVs in 246 real-world Java projects and their corresponding expert-verified CVSS metrics on NVD. Distributions of curated CVSS metrics are illustrated in Fig.~\mbox{\ref{fig:cvss_distribution}}.
The details of the number of commits and projects retained in each filtering step are also given in Table~\ref{tab:filtering_details}. Note that some commits and projects were removed during the tracing of VCCs from VFCs due to the issues coined as ghost commits studied by Rezk et al.~\cite{rezk2021ghost}. We did not remove large VCCs (with more than 100 files and 10k lines) as we found several VCCs were large initial/first commits. Our observations agreed with the findings of Meneely et al.~\cite{meneely2013patch}.

\begin{table}[t]
\fontsize{6.9}{7.9}\selectfont

  \centering
  \caption{The number of commits and projects after each filtering step.}
 \begin{tabular}{llll}
    \hline
    \textbf{No.} & \textbf{Filtering step} & \textbf{No. of commits} & \textbf{No. of projects} \\
    \hline
    1 & All unfiltered VFCs & 13,310 & 2,864 \\
    2 & Removing duplicate VFCs & 9,989 & 2,864 \\
    3 & Removing non-Java VFCs  & 1,607 & 361 \\
    4 & \makecell[l]{Removing VFCs with more than\\100 files \& 10k lines} & 1,602 & 358 \\
    5 & \makecell[l]{Tracing VCCs from VFCs using\\the SZZ algorithm} & 3,742 & 342 \\
    6 & \makecell[l]{Removing VCCs with null\\characteristics (CVSS values)} & 2,271 & 246 \\
    7 & Removing duplicate VCCs & 1,229 & 246 \\
    \hline
    \end{tabular}
  \label{tab:filtering_details}
\end{table}

\noindent \textbf{Manual VCC validation}. To validate our curated VCCs, we randomly selected 293 samples, i.e., 95\% confidence level and 5\% error~\mbox{\cite{cochran2007sampling}}, for two authors to independently examine. The manual VCC validation was considerably labor-intensive, which took approximately 120 man-hours.
The Cohen's kappa ($\kappa$) inter-rater reliability score~\mbox{\cite{mchugh2012interrater}} was 0.83, i.e., ``almost perfect'' agreement~\mbox{\cite{hata20199}}.
We also involved the third author in the discussion to resolve disagreements. Our validation found that 85\% of the VCCs were valid. In fact, the SZZ algorithm is imperfect~\mbox{\cite{fan2019impact}}, but we assert that it is nearly impossible to obtain near 100\% accuracy without exhaustive manual validation. Specifically, the main source of incorrectly identified VCCs in our dataset was that some files in VFCs were used to update version/documentation or address another issue instead of fixing an SV. One such false positive VCC was the commit \textit{87c89f0} in the \textit{jspwiki} project that last modified the build version in the corresponding VFC.

\begin{figure}[t]
    \centering
    \includegraphics[width=\columnwidth,keepaspectratio]{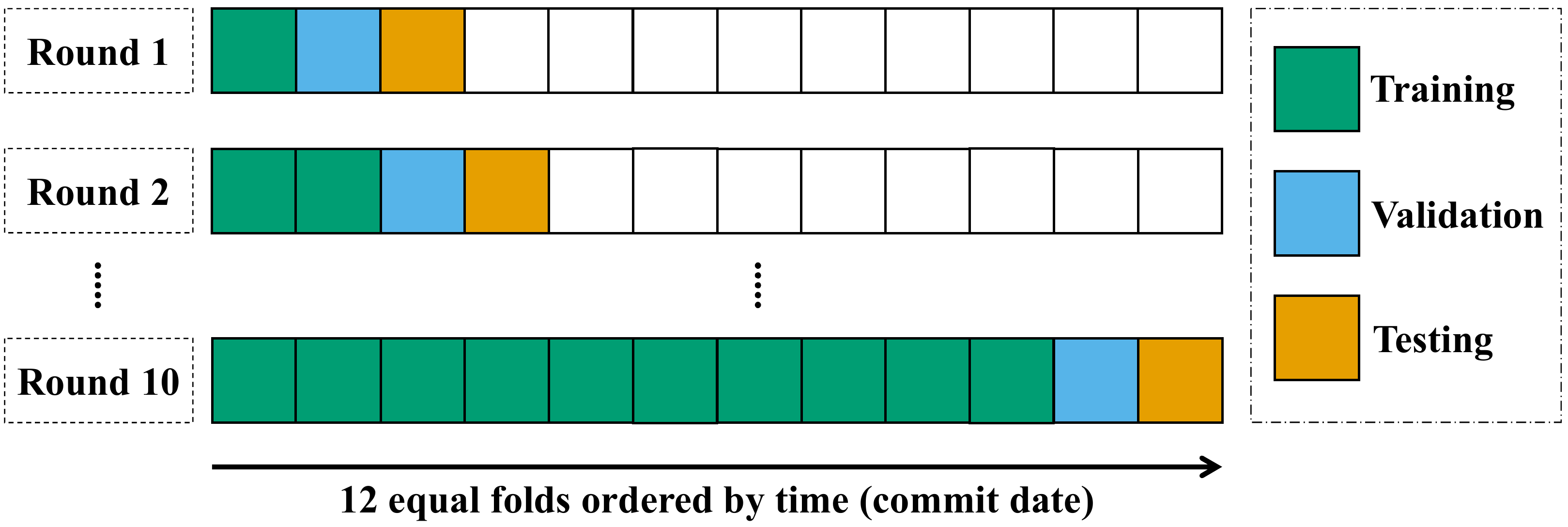}
    \caption{Time-based splits for training, validating \& testing.}
    \label{fig:timebased_splits}
\end{figure}

\noindent \textbf{Data splitting}. We adopted \textit{time-based} splits~\mbox{\cite{falessi2020need}} for training, validating and testing the models to closely represent real-world scenarios where incoming/future unseen data is \textit{not} present during training~\mbox{\cite{mcintosh2017fix,jimenez2019importance}}. We trained, validated and tested the models in 10 rounds using 12 equal folds split based on commit dates (see Fig.~\mbox{\ref{fig:timebased_splits}}). Specifically, in round $i$, folds $1 \rightarrow i$, $i+1$ and $i+2$ were used for training, validation and testing, respectively. We chose an optimal model with the highest average \textit{validation} performance and then reported its respective average testing performance over 10 rounds, which helped avoid unstable results of a single testing set~\mbox{\cite{raschka2018model}}.

\subsection{Evaluation Metrics}
\label{subsec:evaluation_metrics}

To evaluate the performance of automated commit-level SV assessment, we utilized the F1-Score and Matthews Correlation Coefficient (MCC) metrics that have been commonly used in the literature (e.g.,~\mbox{\cite{han2017learning,spanos2018multi,jimenez2019importance}}). These two metrics are suitable for the imbalanced classes~\mbox{\cite{luque2019impact}} in our data (see Fig.~\mbox{\ref{fig:cvss_distribution}}). F1-Score has a range from 0 to 1, while MCC takes values from –1 to 1, where 1 is the best value for both metrics. MCC was used to select optimal models since MCC explicitly considers all classes~\mbox{\cite{luque2019impact}}.
To evaluate the tasks with more than two classes, we used macro F1-Score~\mbox{\cite{spanos2018multi}} and the multi-class version of MCC~\mbox{\cite{gorodkin2004comparing}}. MCC of the multi-task DeepCVA model was the average MCC of seven constituent tasks. Note that MCC is not directly proportional to F1-score.

\subsection{Hyperparameter and Training Settings of DeepCVA}
\label{subsec:tuning_deepcva}

\noindent \textbf{Hyperparameter settings}. We used the average \textit{validation} MCC to select optimal hyperparameters for DeepCVA's components. We also ran DeepCVA 10 times each round to reduce the impact of random initialization on model performance. We first chose 1024 for the input length of the pre-/post-change hunks/context (see Fig.~\mbox{\ref{fig:workflow_deepcva}}), which has been commonly used in the literature (e.g.,~\mbox{\cite{devlin2018bert,radford2019language}}).
Using a shorter input length would likely miss many code tokens, while a longer length would significantly increase the model complexity and training time.
Shorter commits were padded with zeros, and longer ones were truncated to ensure the same input size for parallelization with GPU~\mbox{\cite{hoang2019deepjit,hoang2019patchnet}}. We built a vocabulary of 10k most frequent code tokens in the Input Embedding layer as suggested by~\mbox{\cite{pradel2018deepbugs}}.
Note that using 20k-sized vocabulary only raised the performance by 2\%, yet increased the model complexity by nearly two times.
We selected an input embedding size of 300, i.e., a standard and usually high limit value for many embedding models (e.g.,~\mbox{\cite{mikolov2013distributed,bojanowski2017enriching}}), and we randomly initialized embedding vectors~\mbox{\cite{hoang2019deepjit,kim2014convolutional}}. For the number of filters of the three-way CNN as well as the hidden units of the GRU, Attention and Task-specific blocks, we tried \{32, 64, 128\}, similar to~\mbox{\cite{han2017learning}}. We picked 128 as it had at least 5\% better validation performance than 32 and 64.

\noindent \textbf{Training settings}.
We used the Adam algorithm~\mbox{\cite{kingma2014adam}}, the state-of-the-art stochastic gradient descent method, for training DeepCVA end-to-end with a learning rate of 0.001 and a batch size of 32 as recommended by Hoang et al.~\mbox{\cite{hoang2019deepjit}}. To increase the training stability, we employed Dropout~\mbox{\cite{srivastava2014dropout}} with a dropout rate of 0.2 and Batch Normalization~\mbox{\cite{ioffe2015batch}} between layers. We trained DeepCVA for 50 epochs, and we would stop training if the \textit{validation} MCC did not change in the last five epochs to avoid overfitting~\mbox{\cite{hoang2019deepjit,hoang2019patchnet}}.

\begin{table*}[t]
\fontsize{6.9}{7.9}\selectfont
  \centering
  \caption{Testing performance of DeepCVA and baseline models. \textbf{Notes}: Optimal classifiers of S-CVA/X-CVA and optimal cluster no. (\textit{k}) of U-CVA are in parentheses. BoW, W2V and SM are Bag-of-Words, Word2vec and software metrics, respectively. The best performance of DeepCVA is from the run with the highest MCC in each round. Best row-wise values are in grey.}
    \begin{tabular}{l|l|ccc|ccc|ccc|c}
    \hline
    \multirowcell{3}[0ex][l]{\textbf{CVSS metric}} & \multirowcell{3}[0ex][l]{\textbf{Evaluation}\\ \textbf{metric}} & \multicolumn{10}{c}{\textbf{Model}}\\
    \cline{3-12}
    & & \multicolumn{3}{c|}{\textbf{S-CVA}} & \multicolumn{3}{c|}{\textbf{X-CVA}} & \multicolumn{3}{c|}{\textbf{U-CVA}} & \multirowcell{2}{\textbf{DeepCVA (Best}\\ \textbf{in parentheses)}} \\
    \cline{3-11}
    & & \textbf{BoW} & \textbf{W2V} & \textbf{SM} & \textbf{BoW} & \textbf{W2V} & \textbf{SM} & \textbf{BoW} & \textbf{W2V} & \textbf{SM} &  \\
    \hline
    \multirowcell{2}[0ex][l]{\textbf{Confidentiality}} & \textbf{F1-Score} & 0.416 & 0.406 & 0.423 & 0.420 & 0.434 & 0.429 & 0.292 & 0.332 & 0.313 & \cellcolor[HTML]{C0C0C0} \textbf{0.436 (0.475)} \\
    \hhline{~*{11}{-}}
    & \textbf{MCC} & \makecell{0.174\\ (LR)} & \makecell{0.239\\ (LGBM)} & \makecell{0.232\\ (XGB)} & \makecell{0.188\\ (LR)} & \makecell{0.241\\ (LR)} & \makecell{0.203\\ (XGB)} & \makecell{0.003\\ (50)} & \makecell{0.092\\ (45)} & \makecell{0.017\\ (50)} & \cellcolor[HTML]{C0C0C0} \textbf{0.268 (0.299)}  \\
    \hline
    \multirowcell{2}[0ex][l]{\textbf{Integrity}} & \textbf{F1-Score} & 0.373 & 0.369 & 0.352 & 0.391 & 0.415 & 0.407 & 0.284 & 0.305 & 0.330 & \cellcolor[HTML]{C0C0C0} \textbf{0.430 (0.458)} \\
    \hhline{~*{11}{-}}
    & \textbf{MCC} & \makecell{0.127\\ (LGBM)} & \makecell{0.176\\ (LGBM)} & \makecell{0.146\\ (RF)} & \makecell{0.114\\ (LGBM)} & \makecell{0.160\\ (LR)} & \makecell{0.128\\ (LGBM)} & \makecell{-0.005\\ (25)} & \makecell{0.091\\ (30)} & \makecell{0.084\\ (25)} & \cellcolor[HTML]{C0C0C0} \textbf{0.250 (0.295)} \\
    \hline
    \multirowcell{2}[0ex][l]{\textbf{Availability}} & \textbf{F1-Score} & 0.381 & 0.389 & 0.384 & 0.424 & 0.422 & 0.406 & 0.254 & 0.332 & 0.238 & \cellcolor[HTML]{C0C0C0} \textbf{0.432 (0.475)} \\
    \hhline{~*{11}{-}}
    & \textbf{MCC} & \makecell{0.182\\ (RF)} & \makecell{0.173\\ (LGBM)} & \makecell{0.126\\ (XGB)} & \makecell{0.187\\ (LR)} & \makecell{0.192\\ (LR)} & \makecell{0.123\\ (XGB)} & \makecell{0.064\\ (10)} & \makecell{0.092\\ (45)} & \makecell{0.016\\ (3)} & \cellcolor[HTML]{C0C0C0} \textbf{0.273 (0.303)} \\
    \hline
    \multirowcell{2}[0ex][l]{\textbf{Access Vector}} & \textbf{F1-Score} & 0.511 & 0.487 & 0.440 & 0.499 & 0.532 & 0.487 & 0.477 & 0.477 & 0.477 & \cellcolor[HTML]{C0C0C0} \textbf{0.554 (0.578)} \\
    \hhline{~*{11}{-}}
    & \textbf{MCC} & \makecell{0.07\\ (XGB)} & \makecell{0.051\\ (LR)} & \makecell{0.018\\ (LR)} & \makecell{0.044\\ (LGBM)} &  \makecell{0.107\\ (LR)} & \makecell{0.012\\ (LGBM)} & \makecell{0.000\\ (9)} & \makecell{0.000\\ (40)} & \makecell{0.000\\ (6)} & \cellcolor[HTML]{C0C0C0} \textbf{0.129 (0.178)} \\
    \hline
    \multirowcell{2}[0ex][l]{\textbf{Access Complexity}} & \textbf{F1-Score} & 0.437 & 0.448 & 0.417 & 0.412 & 0.445 & 0.361 & 0.315 & 0.365 & 0.385 & \cellcolor[HTML]{C0C0C0} \textbf{0.464 (0.475)} \\
    \hhline{~*{11}{-}}
    & \textbf{MCC} & \makecell{0.119\\ (LR)} & \makecell{0.143\\ (XGB)} & \makecell{0.111\\ (LGBM)} & \makecell{0.131\\ (LR)} & \makecell{0.121\\ (XGB)} & \makecell{0.088\\ (SVM)} & \makecell{0.000\\ (4)} & \makecell{0.022\\ (30)} & \makecell{0.119\\ (15)} & \cellcolor[HTML]{C0C0C0} \textbf{0.242 (0.261)} \\
    \hline
    \multirowcell{2}[0ex][l]{\textbf{Authentication}} & \textbf{F1-Score} & 0.601 & 0.584 & 0.593 & 0.541 & 0.618 & 0.586 & 0.458 & 0.526 & 0.492 & \cellcolor[HTML]{C0C0C0} \textbf{0.657 (0.677)} \\
    \hhline{~*{11}{-}}
    & \textbf{MCC} & \makecell{0.258\\ (SVM)} & \makecell{0.264\\ (XGB)} & \makecell{0.268\\ (LGBM)} & \makecell{0.212\\ (RF)} & \makecell{0.282\\ (SVM)} & \makecell{0.208\\ (XGB)} & \makecell{0.062\\ (50)} & \makecell{0.162\\ (30)} & \makecell{0.089\\ (50)} & \cellcolor[HTML]{C0C0C0} \textbf{0.352 (0.388)} \\
    \hline
    \multirowcell{2}[0ex][l]{\textbf{Severity}} & \textbf{F1-Score} & 0.407 & 0.357 & 0.345 & 0.382 & 0.381 & 0.358 & 0.283 & 0.288 & 0.287 & \cellcolor[HTML]{C0C0C0} \textbf{0.424 (0.460)} \\
    \hhline{~*{11}{-}}
    & \textbf{MCC} & \makecell{0.144\\ (LR)} & \makecell{0.153\\ (XGB)} & \makecell{0.057\\ (XGB)} & \makecell{0.130\\ (LR)} & \makecell{0.149\\ (LGBM)} & \makecell{0.058\\ (XGB)} & \makecell{-0.018\\ (4)} & \makecell{0.010\\ (15)} & \makecell{0.026\\ (4)} & \cellcolor[HTML]{C0C0C0} \textbf{0.213 (0.277)} \\
    \hline
    \hline
    \multirowcell{2}[0ex][l]{\textbf{Average}} & \textbf{F1-Score} & 0.447 & 0.434 & 0.422 & 0.438 & 0.464 & 0.433 & 0.338 & 0.375 & 0.360 & \cellcolor[HTML]{C0C0C0} \textbf{0.485 (0.514)} \\
    & \textbf{MCC} & 0.153 & 0.171 & 0.137 & 0.144 & 0.179 & 0.117 & 0.015 & 0.067 & 0.050 & \cellcolor[HTML]{C0C0C0} \textbf{0.247 (0.286)} \\
    \hline
    \end{tabular}%
    \vspace{-3pt}
  \label{tab:baseline_comparison}%
\end{table*}%

\subsection{Baseline Models}
\label{subsec:baselines}

We considered three types of learning-based baselines for automated commit-level SV assessment, as learning-based models can automatically extract relevant SV patterns/features from input data for prediction without relying on pre-defined rules.
The baselines were (\textit{i}) \textbf{S-CVA}: \textbf{S}upervised single-task model using either software metrics or text-based features including Bag-of-Words (BoW or token count) and Word2vec~\mbox{\cite{mikolov2013distributed}}; (\textit{ii}) \textbf{X-CVA}: supervised e\textbf{X}treme multi-class model that performed a single prediction for all seven tasks using the above feature types; and (\textit{iii}) \textbf{U-CVA}: \textbf{U}nsupervised model using $k$-means clustering~\mbox{\cite{lloyd1982least}} with the same features as S-CVA/X-CVA.
\textcolor{black}{Note that there was no existing technique for automating commit-level SV assessment, so we could only compare DeepCVA with the compatible techniques proposed for related tasks, as described hereafter.}

Software metrics (e.g.,~\mbox{\cite{kamei2012large,perl2015vccfinder,yang2017vuldigger}}) and text-based features (BoW/Word2vec) (e.g., \mbox{\cite{sabetta2018practical,zhou2017automated}}) have been widely used for commit-level prediction.
We used 84 software metrics proposed by~\mbox{\cite{kamei2012large,perl2015vccfinder,yang2017vuldigger}} for defect/SV prediction.
Among these metrics, we converted C/C++ keywords into Java ones to match our dataset.
The list of software metrics used in this work can be found at~\mbox{\cite{reproduction_package_ase2021}}.
As in~\mbox{\cite{kamei2012large}}, in each round in Fig.~\mbox{\ref{fig:timebased_splits}}, we also removed correlated software metrics that had a Spearman correlation larger than 0.7 based on the training data of that round to avoid performance degradation, e.g., no. of \textit{stars} vs. \textit{forks} of a project.
For BoW and Word2vec, we adopted the same vocabulary size of 10k to extract features from four inputs described in Fig.~\mbox{\ref{fig:workflow_deepcva}}, as in DeepCVA. Feature vectors of all inputs were concatenated into a single vector.
For Word2vec, we averaged the vectors of all tokens in an input to generate its feature vector, which has been shown to be a strong baseline~\mbox{\cite{shen2018baseline}}.
Like DeepCVA, we also used an embedding size of 300 for each Word2vec token.

Using these feature types, S-CVA trained a separate supervised model for each CVSS task, while X-CVA used a single multi-class model to predict all seven tasks simultaneously. X-CVA worked by concatenating all seven CVSS metrics into a single label.
To extract the results of the individual tasks for X-CVA, we checked whether the ground-truth label of each task was in the concatenated model output. For S-CVA and X-CVA, we applied six popular classifiers: Logistic Regression (LR), Support Vector Machine (SVM), K-Nearest Neighbors (KNN), Random Forest (RF), XGBoost (XGB)~\mbox{\cite{chen2016xgboost}} and Light Gradient Boosting Machine (LGBM)~\mbox{\cite{ke2017lightgbm}}. These classifiers have been used for SV assessment based on SV reports~\mbox{\cite{spanos2018multi,le2019automated}}. The hyperparameters for tuning these classifiers were \textit{regularization}: \{l1, l2\}; \textit{regularization coefficient}: \{0.01, 0.1, 1, 10, 100\} for LR and \{0.01, 0.1, 1, 10, 100, 1,000, 10,000\} for SVM; \textit{no. of neighbors}: \{11, 31, 51\}, \textit{distance norm}: \{1, 2\} and \textit{distance weight}: \{uniform, distance\} for KNN; \textit{no. of estimators}: \{100, 300, 500\}, \textit{max. depth}: \{3, 5, 7, 9, unlimited\}, \textit{max. no. of leaf nodes}: \{100, 200, 300, unlimited\} for RF, XGB and LGBM. These hyperparameters have been adapted from relevant studies~\mbox{\cite{spanos2018multi,le2019automated,le2020puminer}}.

Unlike S-CVA and X-CVA, U-CVA did not require CVSS labels to operate; as thus, U-CVA required less human effort than S-CVA and X-CVA. We tuned U-CVA for each task with the following no. of clusters ($k$): \{2, 3, 4, 5, 6, 7, 8, 9, 10, 15, 20, 25, 30, 35, 40, 45, 50\}. To assess a new commit with U-CVA, we found the cluster with the smallest Euclidean distance to that commit and assigned it the most frequent class of each task in the selected cluster.

\section{Research Questions and Experimental Results}
\label{sec:results}

\subsection{\textbf{RQ1}: How does DeepCVA Perform Compared to Baseline Models for Commit-level SV Assessment?}
\label{subsec:rq1_results}

\noindent \textbf{Motivation}. We posit the need for commit-level Software Vulnerability (SV) assessment tasks based on seven CVSS metrics. Such tasks help developers to understand the exploitability and impacts of SVs as early as they are introduced in a system and devise remediation plans accordingly. RQ1 evaluates our DeepCVA for this new and important task.

\noindent \textbf{Method}. We compared the effectiveness of our DeepCVA model with the S-CVA, X-CVA and U-CVA baselines (see section~\mbox{\ref{subsec:baselines}}) on the \textit{testing} sets.
We trained, validated and tested the models using the time-based splits, as described in section~\mbox{\ref{subsec:datasets}}.
Because of the inherent randomness of GPU-based implementation of DeepCVA,\footnote{https://keras.io/getting\_started/faq/\#how-can-i-obtain-reproducible-results-using-keras-during-development} we ran DeepCVA 10 times in each round and then averaged its performance. The baselines were not affected by this issue as they did not use GPU. For DeepCVA, we used the hyperparameter/training settings in section~\mbox{\ref{subsec:tuning_deepcva}}. For each type of baseline, we used grid search on the hyperparameters given in section~\mbox{\ref{subsec:baselines}} to find the optimal model with the highest \textit{validation} MCC (see section~\mbox{\ref{subsec:evaluation_metrics}}).

\noindent \textbf{Results}.
\textit{\textbf{DeepCVA outperformed all baselines}}\footnote{MCC values of \textit{random} and \textit{most-frequent-class} baselines were all $<$ 0.01.} \textit{\textbf{(X-CVA, S-CVA and U-CVA) in terms of both MCC and F1-Score}}\footnote{Precision (0.533)/Recall (0.445) of DeepCVA were $>$ than all baselines.}
\textit{\textbf{for all seven tasks (see Table~\mbox{\ref{tab:baseline_comparison}}).}}
DeepCVA got average and best MCC values of 0.247 and 0.286, i.e., 38\% and 59.8\% better than the second-best baseline (X-CVA with Word2vec features), respectively.
Task-wise, DeepCVA had 11.2\%, 42\%, 42.2\%, 20.6\%, 69.2\%, 24.8\% and 39.2\% higher MCC than the best respective baseline models for Confidentiality, Integrity, Availability, Access Vector, Access Complexity, Authentication and Severity tasks, respectively. Notably, the best DeepCVA model achieved stronger performance than all baselines with MCC percentage gaps from 24.1\% (Confidentiality) to 82.5\% (Access Complexity).
The average and task-wise F1-Score values of DeepCVA also beat those of the best baseline (X-CVA with Word2vec features) by substantial margins.
We found that DeepCVA significantly outperformed the best baseline models in terms of both MCC and F1-score averaging across all seven tasks, confirmed with p-values $<$ 0.01 using the non-parametric Wilcoxon signed-rank tests~\mbox{\cite{wilcoxon1992individual}}.
These results show the effectiveness of the novel design of DeepCVA.

An example to qualitatively demonstrate the effectiveness of DeepCVA is the VCC \textit{ff655ba} in the \textit{Apache xerces2-j} project, in which a hashing algorithm was added. This algorithm was later found vulnerable to hashing collision that could be exploited with timing attacks in the fixing commit \textit{992b5d9}. This SV was caused by the order of items being added to the hash table in the \code{put(String key, int value)} function. Such an order could not be easily captured by baseline models whose features did not consider the sequential nature of code (i.e., BoW, Word2vec and software metrics)~\cite{le2020deep}. More details about the contributions of different components to the overall performance of DeepCVA are covered in section~\mbox{\ref{subsec:rq2_results}}.

Regarding the baselines, the average MCC value (0.147) of X-CVA was on par with that (0.154) of S-CVA. This result reinforces the benefits of leveraging the common attributes among seven CVSS metrics to develop effective commit-level SV assessment models. However, X-CVA was still not as strong as DeepCVA mainly because of its much lower training data utilization per output. For X-CVA, there was an average of 39 output combinations of CVSS metrics in the training folds, i.e., 31 commits per output. In contrast, DeepCVA had 13.2 times more data per output as there were at most three classes for each task (see Fig.~\mbox{\ref{fig:cvss_distribution}}). Finally, we found supervised learning (S-CVA, X-CVA and DeepCVA) to be at least 74.6\% more effective than the unsupervised approach (U-CVA). This result demonstrates the usefulness of using CVSS metrics to guide the extraction of commit features.

\begin{figure*}[t]
    \centering
    \includegraphics[width=\textwidth,keepaspectratio]{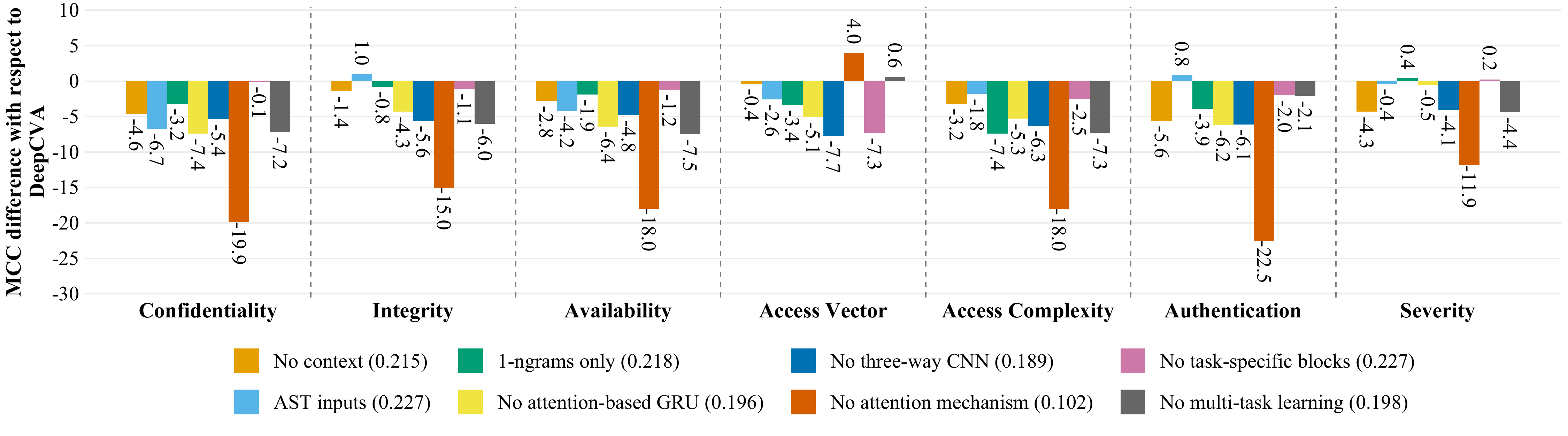}
    \caption{Differences of testing MCC (multiplied by 100 for readability) of the model variants compared to the proposed DeepCVA in section~\mbox{\ref{sec:deepcva_method}}. \textbf{Note}: The average MCC values (without multiplying by 100) of the model variants are in parentheses.}
    \label{fig:deepcva_components}
\end{figure*}

\subsection{\textbf{RQ2}: What are the Contributions of the Main Components in DeepCVA to Model Performance?}
\label{subsec:rq2_results}

\noindent \textbf{Motivation}. We have shown in RQ1 that DeepCVA significantly outperformed all the baselines for seven commit-level SV assessment tasks. RQ2 aims to give insights into the contributions of the key components to such a strong performance of DeepCVA. Such insights can help researchers and practitioners to build effective SV assessment models.

\noindent \textbf{Method}. We evaluated the performance contributions of the main components of DeepCVA: (\textit{i}) Closest Enclosing Scope (CES) of code changes, (\textit{ii}) CNN filter size, (\textit{iii}) Three-way CNN, (\textit{iv}) Attention-based GRU, (\textit{v}) Attention mechanism, (\textit{vii}) Task-specific blocks and (\textit{vi}) Multi-task learning. For each component, we first removed it from DeepCVA, retrained the model variant and reported its \textit{testing} result. When we removed Attention-based GRU, we used max-pooling~\mbox{\cite{hoang2019deepjit,kim2014convolutional}} after the three-way CNN to generate the commit vector. When we removed Multi-task learning, we trained a separate model for each of the seven CVSS metrics. We also investigated an Abstract Syntax Tree (AST) variant of DeepCVA, in which we complemented input code tokens with their syntax (e.g., \code{int a = 1} is a \code{VariableDeclarationStatement}, where \code{a} is an \code{Identifier} and \code{1} is a \code{NumberLiteral}). This AST-based variant explored the usefulness of syntactical information for commit-level SV assessment. We extracted the nodes in an AST that contained code changes and their CES. If more than two nodes contained the code of interest, we chose the one at a lower depth in the AST. We then flattened the nodes with depth-first traversal for feature extraction~\mbox{\cite{lin2018cross}}.

\noindent \textbf{Results}. \textit{\textbf{As depicted in Fig.~\mbox{\ref{fig:deepcva_components}}, the main components}}\footnote{\label{fn:att_mechanism_results}We excluded the DeepCVA variant with no attention mechanism as its performance was abnormally low, affecting the overall trend of other variants.} \textit{\textbf{uplifted the average MCC of DeepCVA by 25.9\% for seven tasks}}. Note that 7/8 model variants (except the model with no attention mechanism) outperformed the best baseline model from RQ1. These results were confirmed with p-values $<$ 0.01 using Wilcoxon signed-rank tests~\mbox{\cite{wilcoxon1992individual}}. Specifically, the components\textsuperscript{\ref{fn:att_mechanism_results}} of DeepCVA increased the MCC by 25.3\%, 20.8\%, 21.5\%, 35.8\%, 35.5\%, 18.9\% and 23.6\% for Confidentiality, Integrity, Availability, Access Vector, Access Complexity, Authentication and Severity, respectively.

For the inputs, using the Smallest Enclosing Scope (CES) of code changes resulted in a 14.8\% increase in MCC compared to using hunks only, while using AST inputs had 8.8\% lower performance.
This finding suggests that code context is important for assessing SVs in commits. In contrast, syntactical information is not as necessary since code structure can be implicitly captured by code tokens and their sequential order using our AC-GRU.

The key components of the AC-GRU feature extractor boosted the performance by 13.2\% (3-grams vs. 1-grams), 25.6\% (Attention-based GRU), 30.2\% (Three-way CNN) and 142\% (Attention). Note that DeepCVA surpassed the state-of-the-art 3-gram~\mbox{\cite{han2017learning}} and 1-gram~\mbox{\cite{hoang2019deepjit}} CNN-only architectures for (commit-level) SV/defect prediction. These results show the importance of combining the (1,3,5)-gram three-way CNN with attention-based GRUs rather than using them individually. We also found that 1-5 grams did not significantly increase the performance (p-value = 0.186), confirming our decision in section~\mbox{\ref{subsec:deep_acgru}} to only use 1,3,5-sized filters.

For the prediction layers, we raised 8.8\% and 24.4\% MCC of DeepCVA with Task-specific blocks and Multi-task learning, respectively. Multi-task DeepCVA took 8,988 s (2.5 hours) and 25.7 s to train/validate and test in 10 rounds $\times$ 10 runs, which were 6.3 and 6.2 times faster compared to those of seven single-task DeepCVA models, respectively. DeepCVA was only 11.3\% and 12.7\% slower in training/validating and testing than one single-task model on average, respectively. These values highlight the efficiency of training and maintaining the multi-task DeepCVA model. Finally, obtaining Severity using the CVSS formula~\mbox{\cite{spanos2018multi}} from the predicted values of the other six metrics dropped MCC by 17.4\% for this task. This result supports predicting Severity directly from commit data.

\subsection{\textbf{RQ3}: What are the Effects of Rebalancing Techniques on Model Performance?}
\label{subsec:rq3_results}

\noindent \textbf{Motivation}. Recent studies (e.g.,~\mbox{\cite{tantithamthavorn2018impact,li2019comparative}}) have shown that rebalancing techniques (i.e., equalizing the class distributions in the training set) can improve model effectiveness for defect/SV prediction. However, these rebalancing techniques can only be applied to single-task models, not multi-task ones. The reason is that each task has a unique class distribution (see Fig.~\mbox{\ref{fig:cvss_distribution}}), and thus balancing class distribution of one task will not balance classes of the others. RQ3 is important to test whether multi-task DeepCVA still outperforms single-task baselines in RQ1/RQ2 using rebalancing techniques.

\noindent \textbf{Method}. We compared the \textit{testing} performance of multi-task DeepCVA with baselines in RQ1/RQ2 using two popular oversampling techniques~\mbox{\cite{tantithamthavorn2018impact}}: \textit{Random OverSampling} (\textit{ROS}) and \textit{SMOTE}~\mbox{\cite{chawla2002smote}}. ROS randomly duplicates the existing samples of minority classes, while SMOTE randomly generates synthetic samples between the existing minority-class samples and their nearest neighbor(s) based on Euclidean distance. We did not consider undersampling, as such models performed poorly because of some very small minority classes (e.g., \textit{Low} Access Complexity had only 14 samples). We applied ROS and SMOTE to \textit{only the training set} and then optimized all baseline models again. Like~\mbox{\cite{tantithamthavorn2018impact}}, we also tuned SMOTE using grid search with different values of nearest neighbors: \{1, 5, 10, 15, 20\}. We could not apply SMOTE to single-task DeepCVA as features were trained end-to-end and unavailable prior training for finding nearest neighbors. We also did not apply SMOTE to X-CVA as there was always a single-sample class in each round, producing no nearest neighbor.

\noindent \textbf{Results}. \textit{\textbf{ROS and SMOTE increased the average performance (MCC) of 3/4 baselines except X-CVA (see Table~\mbox{\ref{tab:rebalancing_results}}). However, the average MCC of our multi-task DeepCVA was still 14.4\% higher than that of the best oversampling-augmented baseline (single-task DeepCVA with ROS)}}. Overall, MCC increased by 8\%, 6.9\% and 9.1\% for S-CVA (ROS), S-CVA (SMOTE) and single-task DeepCVA (ROS), respectively. These improvements were confirmed significant with p-values $<$ 0.01 using Wilcoxon signed-rank tests~\mbox{\cite{wilcoxon1992individual}}. We did not report oversampling results of U-CVA as they were still much worse compared to others. We found single-task DeepCVA benefited the most from oversampling, probably since Deep Learning usually performs better with more data~\mbox{\cite{zheng2020impact}}. In contrast, oversampling did not improve X-CVA as oversampling did not generate as many samples for X-CVA per class as for S-CVA (i.e., X-CVA had 13 times, on average, more classes than S-CVA). These results further strengthen the effectiveness and efficiency of multi-task learning of DeepCVA for commit-level SV assessment even without the overheads of rebalancing/oversampling data.

\begin{table}[t]
\fontsize{6.9}{7.9}\selectfont
  \centering
  \caption{Testing performance (MCC) of optimal baselines using oversampling techniques and multi-task DeepCVA. \textbf{Note}: \textsuperscript{\textdagger}denotes that the oversampled models outperformed the non-oversampled one reported in RQ1/RQ2.}
    \begin{tabular}{l|ccccc}
    \hline
    \textbf{CVSS Task} & \makecell[c]{\textbf{S-CVA}\\ \textbf{(ROS)}} & \makecell[c]{\textbf{S-CVA}\\ \textbf{(SMOTE)}} & \makecell[c]{\textbf{X-CVA}\\ \textbf{(ROS)}} & \makecell[c]{\textbf{Single-task}\\ \textbf{DeepCVA}\\ \textbf{(ROS)}} & \makecell[c]{\textbf{Multi-task}\\ \textbf{DeepCVA}}\\
    \hline
    \textbf{Confidentiality} & 0.220 & 0.203 & 0.185 & 0.250\textsuperscript{\textdagger} & \cellcolor[HTML]{C0C0C0} \textbf{0.268} \\
    \textbf{Integrity} & 0.174 & 0.168 & 0.179\textsuperscript{\textdagger} & 0.206\textsuperscript{\textdagger} & \cellcolor[HTML]{C0C0C0} \textbf{0.250} \\
    \textbf{Availability} & 0.195\textsuperscript{\textdagger} & 0.187\textsuperscript{\textdagger} & 0.182 & 0.209\textsuperscript{\textdagger} & \cellcolor[HTML]{C0C0C0} \textbf{0.273} \\
    \textbf{Access Vector} & 0.115\textsuperscript{\textdagger} & 0.110\textsuperscript{\textdagger} & 0.092 & \cellcolor[HTML]{C0C0C0} \textbf{0.156}\textsuperscript{\textdagger} & 0.129 \\
   \textbf{Access Comp.} & 0.172\textsuperscript{\textdagger} & 0.186\textsuperscript{\textdagger} & 0.144\textsuperscript{\textdagger} & 0.190\textsuperscript{\textdagger} & \cellcolor[HTML]{C0C0C0} \textbf{0.242} \\
    \textbf{Authentication} & 0.325\textsuperscript{\textdagger} & 0.340\textsuperscript{\textdagger} & 0.299\textsuperscript{\textdagger} & 0.318 & \cellcolor[HTML]{C0C0C0} \textbf{0.352} \\
    \textbf{Severity} & 0.132 & 0.124 & 0.141 & 0.186\textsuperscript{\textdagger} & \cellcolor[HTML]{C0C0C0} \textbf{0.213} \\
    \hline
    \hline
    \textbf{Average} & 0.190\textsuperscript{\textdagger} & 0.188\textsuperscript{\textdagger} & 0.175 & 0.216\textsuperscript{\textdagger} & \cellcolor[HTML]{C0C0C0} \textbf{0.247} \\
    \hline
    \end{tabular}%
  \label{tab:rebalancing_results}%
\end{table}%

\section{Discussion}
\label{sec:discussion}

\subsection{DeepCVA and Beyond}

DeepCVA has been shown to be effective for commit-level SV assessment in the three RQs, but our model still has false positives. We analyze several representative patterns of such false positives to help further advance this task and solutions for researchers and practitioners.

Some commits were too complex and large to be assessed correctly. For example, the VCC \textit{015f7ef} in the \textit{Apache Spark} project contained 1,820 additions and 146 deletions across 29 files; whereas, the untrusted deserialization SV occurred in just one line 56 in \code{LauncherConnection.java}. Recent techniques (e.g.,~\mbox{\cite{wattanakriengkrai2020predicting,pascarella2019fine}}) can pinpoint more precise locations (e.g., individual files or lines in commits) of defects.
Such techniques can be adapted to remove irrelevant code in VCCs (i.e., changes that do not introduce or contain SVs). More relevant code potentially gives more fine-grained information for the SV assessment tasks. Note that DeepCVA provides a strong baseline for comparing against fine-grained approaches.

DeepCVA also struggled to predict assessment properties for SVs related to external libraries. For instance, the SV in the commit \textit{015f7ef} above occurs with the \code{ObjectInputStream} class from the \code{java.io} package, which sometimes prevented DeepCVA from correctly assessing an SV. If an SV happens frequently with a package in the training set, (e.g., the XML library of the VCC \textit{bba4bc2} in Fig.~\mbox{\ref{fig:commit_ex}}), DeepCVA still can infer correct CVSS metrics. Pre-trained code models on large corpora~\mbox{\cite{alon2019code2vec,devlin2018bert,hoang2020cc2vec}} along with methods to search/generate code~\mbox{\cite{gu2018deep}} and documentation~\mbox{\cite{hu2018deep}} as well as (SV-related) information from developer Q\&A forums~\cite{le2021large} can be investigated to provide enriched context of external libraries, which would support more reliable commit-level SV assessment with DeepCVA.

We also observed that DeepCVA, alongside the considered baseline models, performed significantly worse, in terms of MCC, for Access Vector compared to the remaining tasks. We speculate that the main reason for such low performance is due to Access Vector containing the most significant class imbalance among the tasks, as shown in Fig.~\ref{fig:cvss_distribution}. For single-task models, we found that using data rebalancing techniques such as ROS or SMOTE can help improve the performance, as demonstrated in RQ3 (see section~\ref{subsec:rq3_results}). However, it is still unclear how to apply the current data rebalancing techniques for multi-task learning models such as DeepCVA. Thus, we suggest that more future work should investigate specific data rebalancing/augmentation to address such imbalanced data in the context of multi-task learning.

\subsection{Threats to Validity}

The first threat is the collection of VCCs. We followed the practices in the literature to reduce the false positives of the SZZ algorithm. We further mitigated this threat by performing independent manual validation with three of the authors.

Another concern is the potential suboptimal tuning of baselines and DeepCVA. However, it is impossible to try the entire hyperparameter space within a reasonable amount of time. For the baselines, we lessened this threat by using a wide range of hyperparameters of baseline models from the previous studies to reoptimize these models from scratch on our data. For DeepCVA, we adapted the best practices recommended in the relevant literature to our tasks.

The reliability and generalizability of our findings are also potential threats. We ran DeepCVA 10 times to mitigate the experimental randomness. We confirmed our results using non-parametric statistical tests with a confidence level $>$ 99\%. Our results may not generalize to all software projects. However, we reduced this threat by conducting extensive experiments on 200+ real-world projects of different scales and domains.

\section{Related Work}
\label{sec:related_work}

\subsection{Data-driven SV Prediction and Assessment}

Public security databases like NVD and expert-based SV scoring frameworks like CVSS have provided large-scale data to determine different properties of SVs. Bozorgi et al.~\mbox{\cite{bozorgi2010beyond}} pioneered this area by developing a Support Vector Machine model to predict when SVs would be exploited. After that, SV information on NVD has been utilized to infer the types~\mbox{\cite{neuhaus2010security}}, severity level~\mbox{\cite{spanos2017assessment}} and exploitability~\mbox{\cite{bullough2017predicting}} of SVs. Recently, many studies~\mbox{\cite{le2019automated,spanos2018multi,elbaz2020fighting,gong2019joint}} have used data-driven techniques to obtain various CVSS metrics for SV assessment from SV reports on NVD. Other studies~\mbox{\cite{ponta2018beyond, ponta2020detection}} have leveraged code patterns in fixing commits of third-party libraries to assess SVs in such libraries. Our work is fundamentally different from these previous studies since we are the first to investigate the potential of performing assessment of all SV types (not only vulnerable libraries) using commit changes rather than bug/SV reports/fixes. Our approach allows practitioners to realize the exploitability/impacts of SVs in their systems much earlier, e.g., up to 1,000 days before (see section~\mbox{\ref{subsec:cva_need}}), as compared to using bug/SV reports/fixes. Less delay in SV assessment helps practitioners to plan/prioritize SV fixing with fresh design and implementation in their minds. Moreover, we have shown that multi-task learning, i.e., predicting all CVSS metrics simultaneously, can significantly increase the effectiveness and reduce the model development and maintenance efforts in commit-level SV assessment.

\subsection{SV Analytics in Code Changes}

Commit-level prediction (e.g.,~\mbox{\cite{kamei2012large,hoang2019patchnet,yang2015deep}}) has been explored to provide \textit{just-in-time} information for developers about code issues, but such studies mainly focused on generic software defects. However, SV is a special type of defects~\mbox{\cite{camilo2015bugs}} that can threaten the security properties of a software project. Thus, SV requires special treatment~\mbox{\cite{peters2017text}} and domain knowledge~\mbox{\cite{gegick2010identifying}}. Meneely et al.~\mbox{\cite{meneely2013patch}} and Bosu et al.~\mbox{\cite{bosu2014identifying}} conducted in-depth studies on how code and developer metrics affected the introduction and review of VCCs. Besides analyzing the characteristics of VCCs, other studies~\mbox{\cite{perl2015vccfinder,yang2017vuldigger,chen2019large}} also developed commit-level SV detection models that leveraged software and text-based metrics. Different from the previous studies that have detected VCCs, we focus on the assessment of such VCCs. SV assessment is as important as the detection step since assessment metrics help early plan and prioritize remediation for the identified SVs. It is worth noting that the existing SV detection techniques can be used to flag VCCs that would then be assessed by our DeepCVA model.

\section{Conclusions and Future Work}
\label{sec:conclusions}

We introduce DeepCVA, a novel deep multi-task learning model, to tackle a new task of commit-level SV assessment. DeepCVA promptly informs practitioners about the CVSS severity level, exploitability, and impact of SVs in code changes after they are committed, enabling more timely and informed remediation. DeepCVA substantially outperformed many baselines (even the ones enhanced with rebalanced data) for the seven commit-level SV assessment tasks. Notably, multi-task learning utilizing the relationship of assessment tasks helped our model be 24.4\% more effective and 6.3 times more efficient than single-task models.
With the reported performance, DeepCVA realizes the first promising step towards a holistic solution to assessing SVs as early as they appear.

We plan to extend DeepCVA to other programming languages and different SV assessment metrics to make the model even more practical for developers. We also aim to investigate DeepCVA for SV detection and fixing tasks to provide an all-in-one solution for practitioners to detect, assess and fix SVs.

\section*{Acknowledgments}
The work was supported by the Cyber Security Research Centre Limited whose activities are partially funded by the Australian Government's Cooperative Research Centres Programme. This work was supported with supercomputing resources provided by the Phoenix HPC service at the University of Adelaide. We would also like to sincerely thank the members from the Centre for Research on Engineering Software Technologies (CREST), Faheem, Chadni, Bushra, Mubin and Huaming, as well as the anonymous reviewers for the insightful and constructive comments to improve the paper.

\IEEEtriggeratref{73}

\bibliographystyle{IEEEtran}
\bibliography{reference}

\end{document}